\documentclass[fleqn,twoside]{article}
\usepackage{espcrc2,epsfig}

\newcommand{\ba}{\begin{eqnarray}}
\newcommand{\ea}{\end{eqnarray}}
\newcommand{\fig}{Fig.~}
\newcommand{\figs}{Figs.~}
\newcommand{\eq}{Eq.~}
\newcommand{\nr}[1]{(\ref{#1})}

\newcommand{\fr}[2]{{\frac{#1}{#2}\,}}
\newcommand{\msbar}{{\overline{\mbox{\rm MS}}}}

\renewcommand{\)}{\right)}
\newcommand{\lb}{\left\{}
\newcommand{\rb}{\right\}}
\newcommand{\lk}{\left[}
\newcommand{\rk}{\right]}

\newcommand{\6}{\partial}

\newcommand{\rmi}[1]{{\mbox{\scriptsize #1}}}

\newcommand{\bG}{{\beta_{\rm G}}}
\newcommand{\Av}[1]{\left\langle #1 \right\rangle}

\newcommand{\tr}{{\rm Tr}}

\renewcommand{\ss}{\textstyle}
\newcommand{\tinymsbar}{{\overline{\mbox{\tiny\rm{MS}}}}}
\newcommand{\<}{\langle}
\renewcommand{\>}{\rangle}
\newcommand{\aamsbar}{\< \tr {(A_0/g_3)^2} \>_{\tinymsbar}}
\newcommand{\aamspert}{\<\tr {(A_0/g_3)^2} %
\>_{\tinymsbar}^\rmi{pert}}

\title{Four--loop logarithms in 3d gauge + Higgs theory\thanks{
Presented at {\em Lattice 2002} by Y.~Schr\"oder.}}

\author{
K. Kajantie\address[Hki]{Department of Physics,
P.O.Box 64, FIN-00014 University of Helsinki, Finland}, 
M. Laine\address{Theory Division, CERN, CH-1211 Geneva 23,
Switzerland}, 
K. Rummukainen$^{{\rm a},}$\address{NORDITA, Blegdamsvej 17,
DK-2100 Copenhagen \O, Denmark}
and
Y. Schr\"oder\address{Center for Theoretical Physics, MIT, 
Cambridge, MA 02139, USA} }

\begin{document}

\begin{abstract}
\vspace{4mm}
We discuss the logarithmic contributions
to the vacuum energy density 
of the three-dimensional SU(3) + adjoint Higgs
theory in its symmetric phase, and relate them to numerical Monte Carlo
simulations.
We also comment on the implications of these results for perturbative and
non-perturbative determinations of the pressure of finite-temperature
QCD.

\vspace{2mm}

\noindent
HIP-2002-40/TH, CERN-TH/2002-229, 
NORDITA-2002-59 HE, MIT-CTP 3301, hep-lat/0209072

\vspace{-2mm}
\end{abstract}

\maketitle

%
\section{INTRODUCTION}

Strongly interacting quantum field theories, such as QCD, 
require extensive numerical simulations, to obtain a non-perturbative 
understanding from first principles. 
Some regions in parameter space might however be amenable to
analytic methods, which can then be used to obtain a clearer
physical picture as well as an independent check on the Monte 
Carlo (MC) simulations. 
Furthermore, since in practice there are upper limits on 
computing power, one might combine numerical and analytic methods,
to supplement each other and provide for a sufficient tool
in cases where either method alone would fail.

As a concrete example of this interplay, let us study the 
free energy density $f$ (which, in the thermodynamic limit, equals 
the negative pressure $p$) of QCD, at finite temperature $T$ and
vanishing baryon chemical potential,
\ba
 e^{-f(T) \fr{V}T} = \int\!\!{\cal D}\lk A\bar\psi\psi\rk
 e^{-\int_0^{1/T}d\tau \int d^3x {\cal L}_E\lk A\bar\psi\psi\rk},
 \nonumber
\ea 
where ${\cal L}_E$ is the standard QCD Lagrangian.
The free energy can be expected to be a good candidate to witness the 
change of the properties of QCD matter around a critical temperature
$T_c\sim200$ MeV. 
While the low-temperature phase is governed by bound states,
such as mesons, the high-temperature phase should, due to
asymptotic freedom, look more like a gas of free quarks and gluons.

A direct lattice measurement of $f$ can be and has been 
performed, see e.g. \cite{4dlat}. 
The results show the pressure rising sharply around $T_c$, 
to level off at a few times $T_c$. 
At higher temperatures, the direct numerical approach gets 
increasingly harder, since ensuring clean continuum as well 
as thermodynamic limits one is facing a multiscale problem, 
$a\ll\fr1T\ll\fr1{T_c}\approx\fr1{\Lambda_\rmi{$\msbar$}}
\approx 1\mbox{ fm}\ll Na$,
where $a$ and $N$ are the lattice spacing 
and the number of lattice sites,
respectively.

On the other hand, 
the temperature being the only scale in the problem, perturbative
methods are guaranteed to work well at high $T$,
due to asymptotic freedom. 
In fact, at vanishing coupling one 
reaches the ideal-gas limit, 
{\small\ba
p_{\,{\rm ideal}}(T)=\fr{\pi^2 T^4}{45}\(N_c^2\!-\!1+\fr74\,N_cN_f\)\;,
\ea}

\noindent
where $N_c$ and $N_f$ denote the number of colours and flavours,
respectively. 
With decreasing temperature, the value of the effective 
coupling constant $g(T)$ however increases, rendering a perturbative
series \cite{pert} meaningless below some point. 

%
\section{METHOD}

Progress can be made by exploiting the scale hierarchy $\pi T>gT>g^2T$
at high temperatures, enabling one to use the powerful analytic 
method of effectice theories, allowing to reduce numerical simulations
needed for the QCD pressure to a much less demanding three-dimensional 
(3d) bosonic theory \cite{a0condlatt01}. 
The partition function factorizes, and hence the
free energy decomposes into 
$f_{\rm QCD}=f_{\rm hard}+f_{\rm soft}$ \cite{bn}.
The effective theory for the {\em soft}, ${\cal O}(gT)$ modes turns
out to be dimensionally reduced, 
\ba
 e^{-V f_{\rm soft}}=
 \int{\cal D}[A_iA_0] e^{-\int d^3x {\cal L}_\rmi{3d}[A_iA_0]}\;,
\ea
where ${\cal L}_\rmi{3d}$ is a 3d SU(3) + adjoint Higgs theory,
\ba\label{L3d}
 {\cal L}_\rmi{3d} =
 {\ss \fr14 F_{ij}^2 +\fr12\! [D_i,A_0]^2 +\fr12\! m^2_3 A_0^2 
 +\fr14\! \lambda_3 A_0^4}. 
\ea
Its coefficients ($g_3^2$,$m_3^2$,$\lambda_3$) 
are functions of $T$ via perturbative matching \cite{adjoint}, e.g.,
\ba\label{matching}
 y\equiv{\ss\fr{m_3^2}{g_3^4}}\approx{\ss\fr{11}{8\pi^2} 
 \ln\fr{8.086T}                                          
 {\Lambda_\rmi{$\msbar$}}                                
 }\;. 
\ea
Let us note that  
for simplicity, this relation refers to 
the case of pure 4d SU(3), while the inclusion of fermions, 
as well as (small) chemical potentials, is also possible.

For a more detailed account of the setup, we refer
to \cite{a0condlatt01}. In the following we wish to highlight 
the specific role of logarithmic terms in $f_\rmi{soft}$. 

%
\section{3d LATTICE MEASUREMENTS}

\begin{figure}[tb]

\epsfxsize=7.5cm\epsfbox{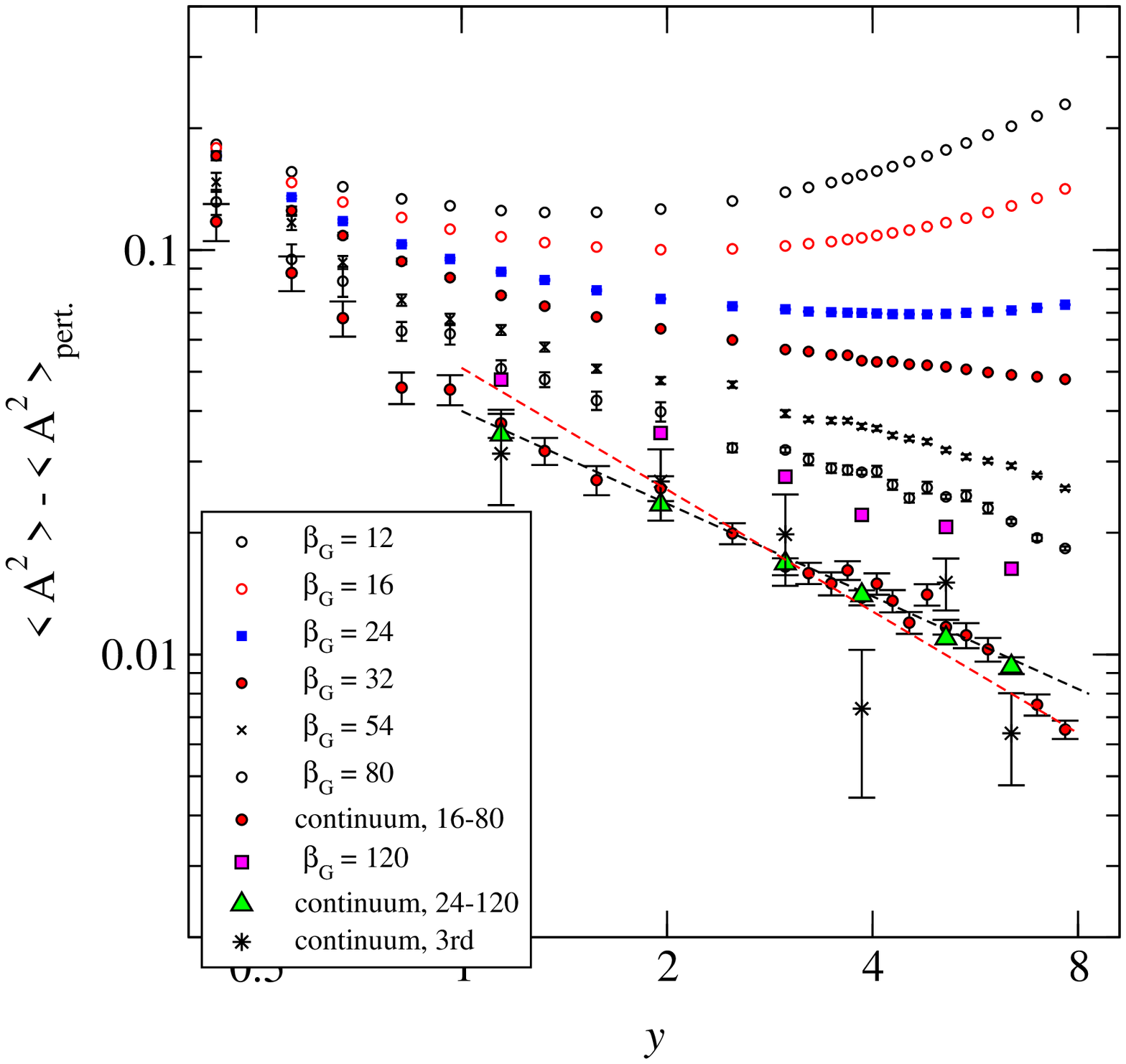}

\vspace{-10mm}

\caption{Lattice results for $\aamsbar-\aamspert$ as a function of $y$, $\bG$.
Various continuum extrapolated values are also shown.} 
\label{fig:A02_1}
\end{figure}

Related to the fact that ${\cal L}_\rmi{3d}$ defines a confining theory, 
it turns out that 
$f_\rmi{soft}$ is perturbatively computable only up to 3-loop level, 
while all higher loop orders contribute at the next level~\cite{nonpert}. 
The parametric form of $f_\rmi{soft}$ can however still be written down
for large $y$, 
\ba
 f_\rmi{soft}(y) & = &  
 f_\rmi{soft,pert}(y) \nonumber \\ 
 & + & \frac{g_3^6}{(4\pi)^4}
 \Bigl(c_1 \ln y + c_2 + {\cal O}(\frac{1}{{y}^{1/2}}) \Bigr).
\ea  

The coefficient $c_1$ here can now be accessed with lattice methods. 
Indeed, $\6_y f_\rmi{soft}$ is related to a gauge-invariant condensate, 
\ba
 \6_y f_\rmi{soft}(y) = g_3^4 \left< \tr A_0^2 \right>.
\ea
If we also subtract the known perturbative part, we see that
\ba
 & & \left< \tr (A_0/g_3)^2 \right> - 
 \left< \tr (A_0/g_3)^2 \right>_\rmi{pert} \nonumber \\
 & & \hspace*{1cm} = 
 \frac{1}{(4\pi)^4}
 \Bigl(c_1 \frac{1}{y} + {\cal O}(\frac{1}{y^{3/2}}) \Bigr). \label{a0y}
\ea

An additional issue which has to be addressed is the 
renormalisation of the condensate. However, due to the 
super-renormalisability of ${\cal L}_\rmi{3d}$, this problem
can be taken care of, with a perturbative 2-loop 
computation~\cite{framework,guy}. Denoting 
\ba
 \bG=\fr6{ag_3^2},
\ea 
the result is, schematically, that a lattice measurement 
can be converted to a continuum 
regularisation (such as $\msbar$) through a relation 
{\small\ba
\Av{\tr A_0^2}_\rmi{$\msbar$} \sim \lim_{\bG\rightarrow\infty}\lb 
\Av{\tr A_0^2}_L\!+\!\bG\!+\!\ln\bG\!+\!1\rb .
\ea}

\begin{figure}[tb]

\epsfxsize=7.5cm\epsfbox{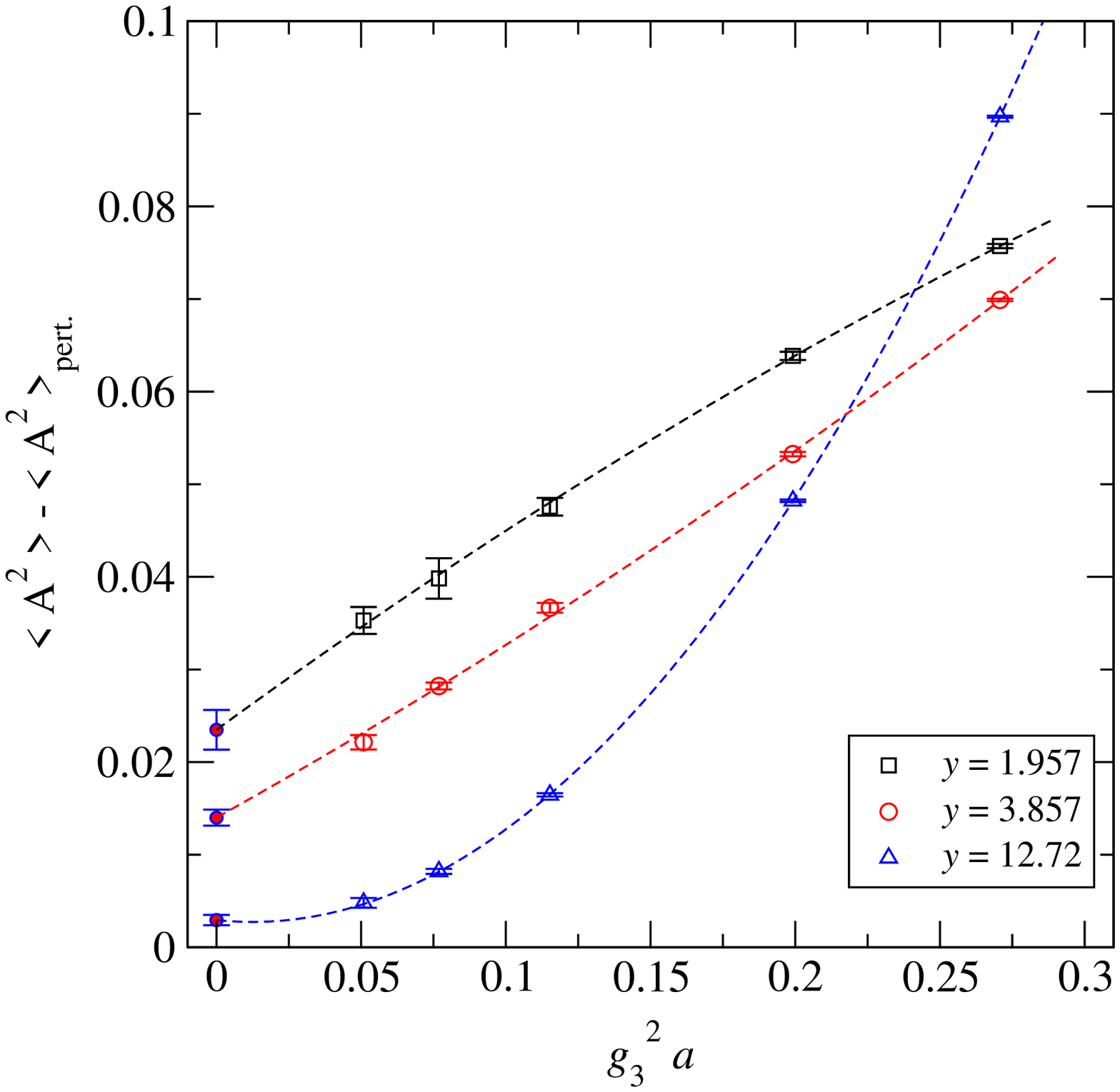}

\vspace{-10mm}

\caption{Examples of continuum extrapolations for $\aamsbar-\aamspert$, 
at a few selected $y$. The fits are polynomial.}  
\label{fig:A02_2}
\end{figure}

\begin{figure}[tb]

\epsfxsize=7.5cm\epsfbox{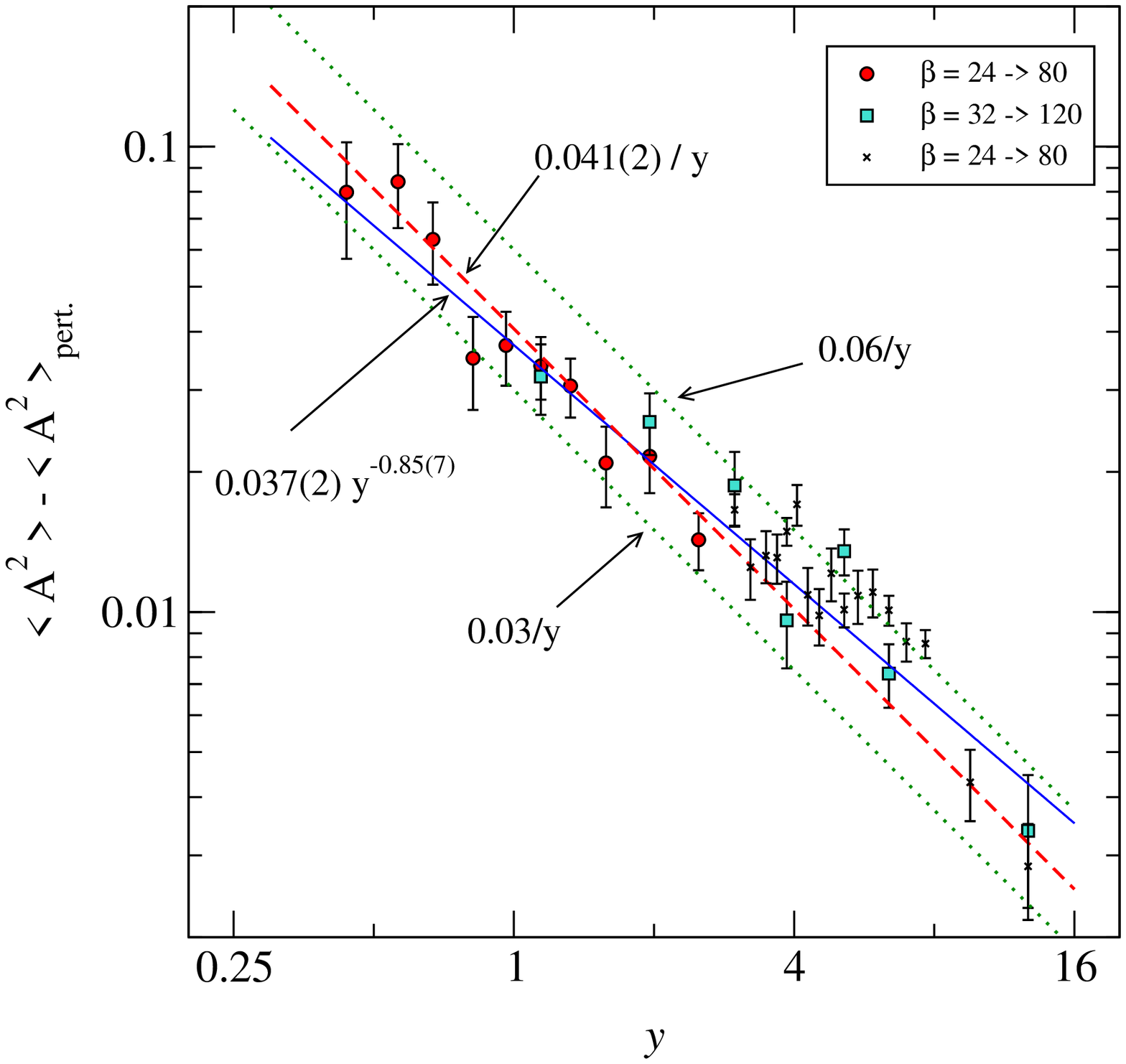}

\vspace{-10mm}

\caption{The continuum extrapolated 
values of $\aamsbar-\aamspert$ as a function 
of $y$, together with various fits. A fit linear in $y^{-1}$ describes 
the data well in a wide range of $y$.}
\label{fig:A02}
\end{figure}

In \figs\ref{fig:A02_1}, \ref{fig:A02_2} we show measurements of the 
condensate with various finite $\bG$, as well as continuum extrapolations. 
The (very preliminary) final result, after the subtraction of the 
3-loop perturbative part, is shown in \fig\ref{fig:A02}.

We find that the data can indeed be well described by the 
functional form in~\eq\nr{a0y}, with what appears to be 
a definite coefficient $c_1$. This clearly calls for 
an analytic evaluation of $c_1$.

Previously~\cite{a0condlatt01}, we have discussed how the 
non-perturbative measurement of $\left< \tr A_0^2 \right>$ 
allows to estimate $f_\rmi{soft}(y)$,  
and correspondingly $f_\rmi{QCD}(T)$,
down to temperatures
of a few times $T_c$. Once $c_1$ is reliably extracted, it
will be interesting to see how well these results can be 
reproduced by keeping in the expression only this single 
logarithm. 

%
\vspace{6mm}
Concluding, we have discussed a method that in principle allows to determine 
the free energy of full QCD from the known analytic limit at high $T$, down 
to a few times $T_c$. 
A small set of perturbative constants 
remains to be determined, but these can already be partly constrained with 
numerical 3d MC data. 

%

\vspace*{3mm}

\noindent{\bf ACKNOWLEDGEMENTS}

\vspace*{3mm}

This work was partly supported by 
the TMR network {\em Finite
Temperature Phase Transitions in Particle Physics}, EU contract no.\ 
FMRX-CT97-0122, 
by the RTN network {\em Supersymmetry and 
the Early Universe}, EU contract no.\ HPRN-CT-2000-00152,
by the Academy of Finland, contract no.\ 77744, 
and by the DOE, under Cooperative Agreement no.\ DF-FC02-94ER40818.


\end{document}